\documentclass[lettersize,journal]{IEEEtran}
\usepackage{xcolor}
\usepackage{amsthm,amssymb,graphicx,multirow,amsmath,color,amsfonts}%,ulem}
\usepackage[update,prepend]{epstopdf}
\usepackage{enumitem}
\usepackage{booktabs}
\usepackage{cite}
\usepackage{amsmath}

\usepackage[latin1]{inputenc}
\usepackage{tikz}
\usetikzlibrary{arrows,calc}		% Optional ticks libraries
\usepackage{bbm} % for \mathbbm{1}
\usepackage{tabulary}
\usepackage{multirow}
\usepackage{comment}
\usepackage{breqn} % for multi line math broken, \begin{dmath}
\usepackage{subcaption}
\usepackage{caption}
\usepackage{placeins} % for don't move figures after section

\usepackage[belowskip=-10pt,aboveskip=0pt]{caption}  % for remove space between fig and captions
\usepackage{graphicx}
\usepackage{caption}
\usepackage{subcaption}

\theoremstyle{definition}
\usepackage{blindtext} % insert figure below eachother

\include{notation}
\allowdisplaybreaks % Allows breaking of eqnarray over multiple pages (avoids unnecessary blanks in the document before eqnarray)

\usepackage[font=small,skip=3pt]{caption}

\begin{document}

	\graphicspath{{./Figures/}}
\title{%Joint coverage and secrecy Analysis  in HAPS-Based Networks Under Jamming coordination
%\vspace{-0.3cm}
   % Jamming Coordination for Secure HAPS-Based Communication: A Joint Coverage and Secrecy Framework
   Secure Wireless Information Transfer and Energy Harvesting in HAPS-Based Networks 
   }

\author{ Khaled Humadi, \textit{Member, IEEE,} Gunes Karabulut-Kurt, \textit{Senior Member, IEEE}, Wessam Ajib, \textit{Senior Member, IEEE}, and Wei-Ping Zhu, \textit{Senior Member, IEEE}
\thanks{
This work  was supported in part by the Fonds de Recherche du Qu\'ebec -  Nature et Technologies (FRQNT), and in part by the Tier 1 Canada Research Chair program. 

Khaled Humadi and  Gunes Karabulut Kurt are with the Poly-Grames Research Center, Department of Electrical Engineering, Polytechnique Montr\'eal, QC, Canada, (email: khaled.humadi@polymtl.ca  and gunes.kurt@polymtl.ca).  Wessam Ajib is with the D\'epartement d'informatique, Universit\'e du Qu\'ebec \`a Montr\'eal, Montr\'eal, QC H2L 2C4, Canada, e-mail: ajib.wessam@uqam.ca. Wei-Ping Zhu is with the Department of Electrical and Computer Engineering, Concordia University, Montreal, QC H3G 1M8, Canada, e-mail: weiping@ece.concordia.ca. }
% \vspace{-0.5cm}
% \vspace{-0.8 cm}
}
    % \author{ \IEEEauthorblockN{ Khaled Humadi\IEEEauthorrefmark{1}, Leila Marandi\IEEEauthorrefmark{1}, Gunes Karabulut-Kurt\IEEEauthorrefmark{1}, Wessam Ajib\IEEEauthorrefmark{2}, Wei-ping Zhu\IEEEauthorrefmark{3}}
    
    % \IEEEauthorblockA{\IEEEauthorrefmark{1} Department of Electrical Engineering, Polytechnique Montreal, Montreal, Canada},\\
    % \IEEEauthorblockA{\IEEEauthorrefmark{2} Department of Computer Sciences, University of Quebec at Montreal, Montreal, Canada},\\
    % \IEEEauthorblockA{\IEEEauthorrefmark{3} Department of Electrical and Computer Engineering, Concordia University, Montreal, Canada.}\\
    % \IEEEauthorblockA{
    %     Email: \{khaled.humadi, leila.marandi, gunes.kurt\}@polymtl.ca, ajib.wessam@uqam.ca, weiping.zhu@concordia.ca.
    % }
    % 

\maketitle
\bstctlcite{IEEEexample:BSTcontrol} % for references:: show author names instead dash

\begin{abstract} 
High-altitude platform station (HAPS) serves as a promising enabler for wide-area connectivity of low-power wireless devices, particularly in remote and underserved regions. However, the strong line-of-sight characteristics of HAPS links increase the risk of eavesdropping, while the limited energy budget of ground devices remains a major operational constraint. In this work, we propose a secure wireless information and energy harvesting framework for HAPS-based networks in the presence of spatially distributed eavesdroppers. The proposed system integrates friendly jamming and power transfer nodes equipped with null-steering capability antennas, such that they not only degrade the reception quality at eavesdroppers but also act as additional radio-frequency energy sources for legitimate users. %For secrecy performance enhancement, jamming nodes within a predetermined zone around each legitimate user employ null-steering to mitigate the interference  toward users while maintaining strong interference in the other directions.
A time-switching wireless information and power transfer   architecture is adopted at the user side. Under this framework, we derive tractable expressions for the joint rate-energy coverage and the average secrecy rate using stochastic geometry tools. Numerical and Monte Carlo results validate the developed analysis and reveal key design trade-offs among the time allocation factor, null-steering-zone radius around each user, and jammer transmit power. In particular, the results show that properly coordinated null-steering jamming can simultaneously support secure communication and adequate wireless power transfer, while an appropriate choice of system parameters, such as time allocation factor and jamming power, is required to balance harvested energy, communication reliability, and secrecy performance.

\end{abstract}

% Geostationary Equatorial Orbit (GEO), Low Earth Orbit (LEO), Unmanned Aerial Vehicle (UAV)
\vspace{-0.3cm}
\section{Introduction} \label{sec:intro}

Recently, high altitude platform stations (HAPS) have gained significant attention as a key component of non-terrestrial networks (NTNs) for beyond-5G and 6G systems \cite{3gpp20183rd}. HAPSs operate in a stratosphere at altitudes of approximately 17-25 km. This enables them to offer an intermediate layer between ground networks and satellites and combines the advantages of both networks. Moreover, HAPS can provide wide coverage and strong line-of-sight (LoS) channels. This makes HAPS highly suitable as a  broadband access in rural and underserved areas \cite{svistunov2025bridging}. 
 In contrast to satellites, the quasi-stationary hovering of HAPS facilitates continuous coverage and effective beamforming, while the reduced altitude dramatically lowers propagation delay and path loss relative to satellites \cite{kurt2021vision}. %Moreover, HAPS can function as aerial base stations or relays to enhance terrestrial communications, hence augmenting network resilience and coverage adaptability. 
 Recent advancements in phased-array antennas and multi-beam transmission enable HAPS to serve a large number of ground users through narrow, high-gain beams, facilitating  spectrum sharing and elevated data rates \cite{javed2024system}.

These inherent properties make HAPS suitable for supporting the Internet of Things (IoT) and low-power devices \cite{giambene2022lora,andreadis2023role,jia2022hierarchical}. Due to  the strong and dominant LoS links and reduced shadowing, reliable communication can be achieved with  low transmit power at the receivers. This is essential for energy-limited  nodes. In addition, the large coverage footprint of HAPS enables connectivity for massive number of IoT devices over a wide and remote area which reduces the need for dense terrestrial infrastructure \cite{svistunov2025bridging}. HAPS can also operate as aerial gateways for data aggregation and processing to facilitate  IoT deployments in applications such as environment monitoring, smart agriculture, and disaster management. Furthermore, by broadcasting RF signals over large coverage areas, HAPS can support wireless power transfer, which enables  IoT devices to harvest energy and operate in a sustainable manner \cite{nauman2023empowering}.

Energy constraints are a significant challenge for low-power devices, including wireless sensors and IoT nodes, functioning in dense and resource-constrained environments \cite{gu2021dynamic,su2020uav}.  Wireless power transfer has garnered significant interest as a potential technique that enables devices to collect energy from ambient RF signals, thereby prolonging their lifespan and reducing the frequent necessity for battery replacement \cite{zhong2022decentralized}.  
Recent studies on SWIPT have focused on enhancing energy efficiency, reliability, and flexibility in next-generation networks.  
The research in \cite{faramarzi2025star_ris_swipt} examines a  simultaneously transmitting and reconfigurable intelligent surface technique for simultaneous wireless information and power transfer (SWIPT). This is obtained by  employing joint optimization of beamforming and power splitting to enhance the system energy efficiency. Cooperative SWIPT with non-orthogonal multiple access has been extensively studied; for example, \cite{ashraf2025swipt_noma} examines a cooperative  system under non-linear energy harvesting, interference, and imperfect successive interference cancellation, deriving essential performance metrics such as outage probability and throughput.  
Mobility-assisted solutions have been suggested to make the system even more adaptable. 
In particular, \cite{feng2025ris_uav_swipt} studies the integration of reconfigurable intelligent surface and  unmanned aerial vehicle (UAV) for  secure SWIPT and jointly optimizes trajectory and resource allocation to improve secrecy performance. Moreover, advanced transmission strategies, such as the integration of successive refinement coding with SWIPT in cooperative non-orthogonal multiple access systems, have been shown to improve both reliability and energy efficiency \cite{wang2026sr_swipt_noma}. Finally, novel hardware designs have been considered, where \cite{huang2025ma_swipt} proposes movable antenna-assisted SWIPT to enhance harvested energy through joint beamforming and antenna position optimization. 
The performance of cellular networks incorporating RF-based wireless power transfer has been  analyzed in several studies through stochastic geometry frameworks. The work in \cite{thanh2018stochastic} investigated how the diversity at the receiver side  can improve   both achievable data rates and harvested energy. 
Meanwhile, \cite{abd2018joint} studied the wireless-powered IoT networks by explicitly modeling the spatial coupling between energy transmitters and IoT devices, where device locations follow a Poisson cluster process centered around IoT gateways.
In \cite{di2016system}, a comprehensive analytical model was introduced to allow system-level evaluation of wireless information and power transfer networks. 
Addressing the impact of void cells, \cite{liu2019energy} develops an analytical model to characterize both uplink and downlink rates, along with the harvested energy in SWIPT-enabled cellular systems. In \cite{deng2021energy}, an energy correlation coefficient is introduced to quantify the influence of correlated energy arrivals under beamforming. 
Furthermore, in \cite {humadi2022simultaneous} SWIPT performance in dense millimeter-wave  networks has been analyzed under user-centric  clustering, highlighting the trade-off between rate and harvested energy. 

Signals transmitted over the wireless medium can be manipulated by unwanted receivers such as eavesdroppers, which directly affect the confidentiality, availability, and integrity of wireless communication systems. Conventional security techniques   have predominantly relied on upper-layer cryptographic techniques such as encryption and authentication protocols \cite{zou2016survey,batewela2025addressing}. Nevertheless, such approaches depend heavily on greedy algorithms, which becomes dramatically complex in large-scale or highly dynamic networks.  
To avoid such computational complexity,  physical layer security (PLS), which exploits channel impairments such as fading, noise, and interference to enhance the wireless system  security, is introduced as an efficient solution, especially in large-scale networks.  \cite{PLS_crypto,Principle_PLS}. 
Many studies have investigated PLS performance in different network deployments. For example, secrecy in heterogeneous and multiple-access systems with randomly distributed eavesdroppers has been analyzed using stochastic geometry tools in \cite{HetNet_noma_Evas}. For millimeter-wave networks, the work in \cite{Secure_mmWave} introduced a cooperative jamming scheme that exploits differences LoS and non-LoS (NLoS) propagation to enhance secrecy.  

However, the investigation of PLS in HAPS-based networks remains relatively limited. Owing to their high-altitude deployment relative to the ground base stations, HAPS platforms enable long-range LoS communication, which increases the vulnerability of transmitted signals to interception over large distances.   In \cite{memarian2025enhancing}, the integration of HAPS and reconfigurable intelligent surfaces techniques has been studied, where joint beamforming and phase optimization are used to enhance the secrecy performance. Hybrid architectures combining free-space optical and terahertz links have been explored in \cite{illi2024secrecy} to exploit multi-HAPS diversity against eavesdropping.  In addition, joint HAPS-UAV communication frameworks are analyzed considering pointing errors and their impact on secrecy metrics in \cite{saber2024security}.

{In this work, we consider a HAPS-based system serving low-power devices such as wireless sensors and IoT nodes. 
To enhance physical layer security, we consider the deployment of friendly jammers equipped with null-steering capabilities that cooperatively operate to enhance the secrecy performance. %In contrast to our previous work \cite{wisee}, where selected jammers are completely deactivated, the proposed scheme in this work enable jammers to direct nulls towards legitimate users while keeping to radiate interference in other directions. 
In contrast to our conference version~\cite{wisee}, which adopts a region-based jammer deactivation (RJD) scheme where all jammers within a protection region of radius $D$ around each user are completely switched off, this work introduces a cooperative null-steering scheme in which these jammers remain active and steer spatial nulls toward the user while continuing to radiate maximum interference in all other directions. %This fundamentally changes the interference structure: rather than removing the nearby jamming power, it is redirected, so that the user is protected while sufficient interference is preserved at the eavesdroppers, thereby enhancing secrecy. 
Furthermore, and unlike~\cite{wisee}, which considers only secure information delivery, this paper incorporates wireless power transfer, whereby each low-power device is powered through energy harvesting, where  collecting energy not only from the HAPS transmission but also from the interference generated by the deployed friendly jammers. %The jamming power that would simply be discarded under deactivation is thus reused as an energy source. 
To the best of the authors' knowledge, this is the first work that considers the integration of HAPS and friendly jammers to provide both secure communication and wireless power transfer for low-power ground receivers. The main contributions of this work are summarized as follows:
\begin{itemize} 
\item We propose a new secure wireless system design in which friendly jamming nodes are not only used to enhance the secrecy performance as in \cite{wisee} but also serve as additional RF energy sources for legitimate users. This provides a unified approach for simultaneous security enhancement and wireless power transfer.%This dual role of the jammers, together with the accompanying wireless power transfer, is not considered in our conference version~\cite{wisee}.
\item In contrast to the jammer deactivation scheme of~\cite{wisee}, we introduce a cooperative null-steering technique in which jamming nodes steer nulls within a carefully selected zone around each legitimate user while remaining active elsewhere. This design reduces the interference toward users while keeping strong interference in the direction of eavesdroppers, which enables a controllable trade-off between energy harvesting and secure communication performance.
\item We derive tractable expressions for the joint rate-energy coverage probability by characterizing the aggregate interference using stochastic geometry tools and the Gil-Pelaez inversion theorem, and then we analyze the average secrecy rate. The joint rate-energy coverage is a new metric introduced by the SWIPT-enabled design of this work, whereas~\cite{wisee} treats coverage and secrecy separately and does not address energy harvesting.
\item Extensive Monte Carlo simulations are conducted to validate the accuracy of the provided analytical results for the joint rate-energy coverage and secrecy rate. In addition, we reveal new design trade-offs, in particular the interplay among the time allocation factor, the null-steering-zone radius, and the jammer power, and provide practical design insights for secure and energy-efficient HAPS-based networks.
\end{itemize}}
The rest of the paper is organized as follows. Section II presents the system models. In Section III, we derive mathematical expressions for the joint rate-energy coverage, while Section IV provides analysis for the average secrecy rate.  Section V presents numerical results that validate the mathematical frameworks and demonstrate the system insights and trade-offs. Finally, Section VI concludes this paper.

%The rest of this article is organized as follows. In Sec. \ref{Sec: System Model}, the system model is outlined and introduced the key parameters. The mathematical expressions are analyzed in Sec. \ref{Sec. Analysis}. Numerical results are presented in Sec. \ref{Sec. Results}. Ultimately,  Sec. \ref{Sec. Conclusion} concludes this article.

% \textit{Notations:} In this context, boldface lowercase letters represent vectors, while boldface uppercase letters represent matrices. $diag(.)$ and $tr(.)$ denote a diagonal and trace matrices. The distribution of a complex Gaussian random vector with a mean of $0$ and a variance of $\delta^{2}$ is denoted as $\mathit{CN}(0; \delta^{2})$. The notations of this paper are summarized in Table (\ref{table:Notation}).

%\vspace{-0.2cm}
\begin{figure*}[t]
    \centering
    \includegraphics[width=\textwidth]{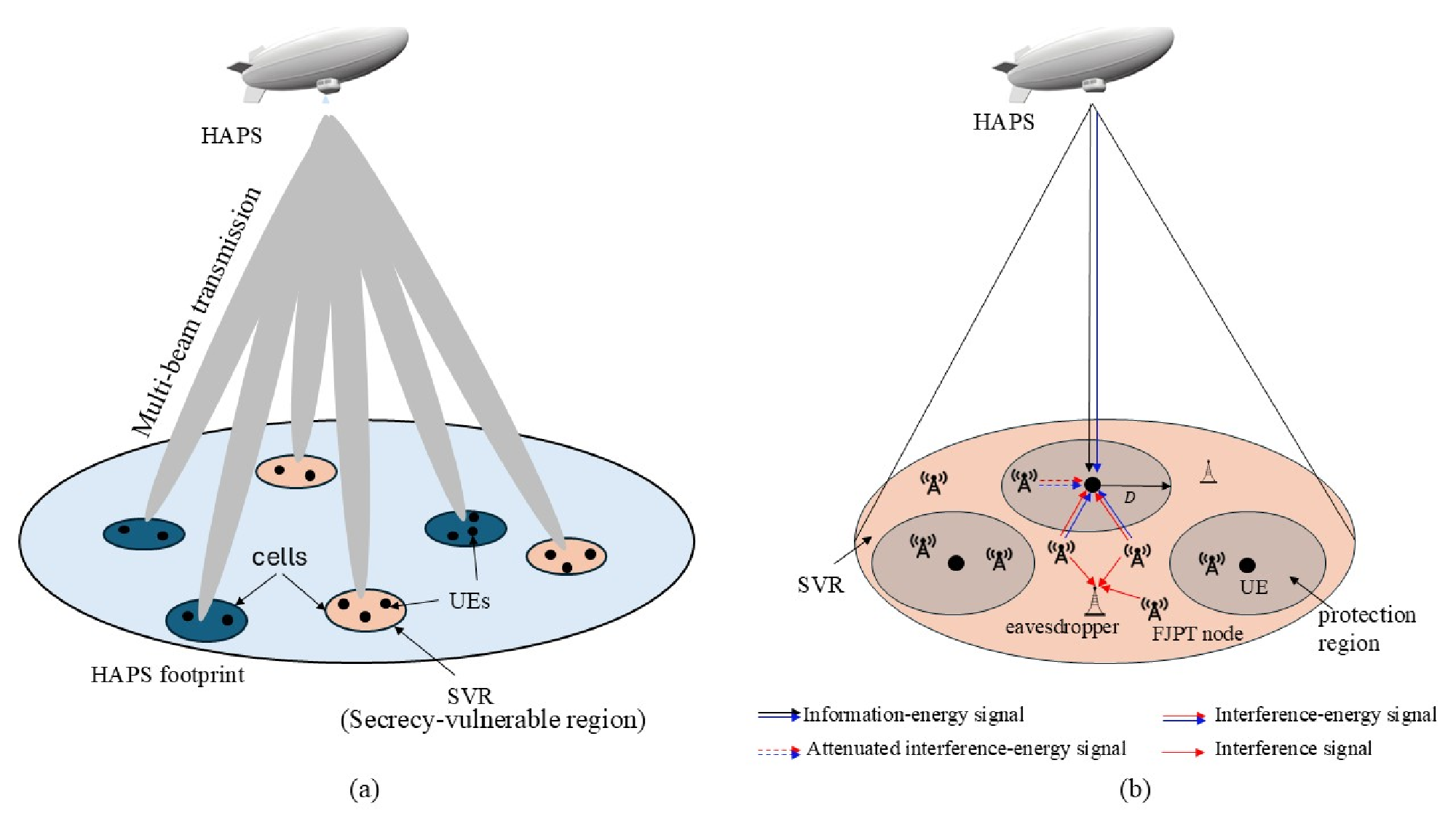}
    \caption{(a) HAPS-based wireless network; (b) secured HAPS-based communication  with wireless power transfer.}
    \label{SMfig}
\end{figure*}
\section{System Model}\label{Sec: System Model}
%\vspace{-0.1cm}
\subsection{Network Model}
For a HAPS serving a given ground coverage area, the footprint is divided into multiple directional beams. Each beam covers a distinct circular region on the ground. Some of these regions are particularly vulnerable to eavesdropping, where stringent security requirements must be enforced. Such regions are referred to as secrecy-vulnerable regions (SVRs), as illustrated in Fig.~\ref{SMfig}a.

In this work, we consider a secure HAPS-based wireless system with simultaneous information and power transfer, where a HAPS located at altitude $a$ transmits confidential information and energy to wirelessly powered user equipments (UEs) in the downlink. The UEs are assumed to be low-power devices without a permanent energy source, relying solely on harvested energy from ambient RF signals. We focus on a single SVR of radius $R_s$, where the wirelessly powered UEs are spatially distributed according to a homogeneous Poisson point process (PPP), denoted by $\Phi_u$ with density $\lambda_u$.

To simultaneously enhance data security and energy harvesting performance, we consider the deployment of friendly jamming and power transfer (FJPT) nodes. These nodes transmit interference in a cooperative manner, which degrades the received signal quality at potential eavesdroppers while serving as an additional RF energy source for legitimate UEs, as shown in Fig.~\ref{SMfig}b. The FJPT nodes and eavesdroppers are modeled as two independent PPPs, denoted by $\Phi_j$ and $\Phi_e$, with densities $\lambda_j$ and $\lambda_e$, respectively. Without loss of generality, we analyze the performance of a reference UE located at the center of the SVR. The FJPT nodes are assumed to be equipped with directional antennas capable of null steering. Furthermore, a cooperative transmission scheme among FJPT nodes is considered. Specifically, all nodes located within a circular region of radius $D$ around each UE steer spatial nulls toward the UE while transmitting with maximum gain in other directions. This region is referred to as the null-steering zone (NSZ). The UE harvests energy from the signals transmitted by the serving HAPS, as well as from the attenuated (or null-steered) signals originating from FJPT nodes within the NSZ, and the interference signals transmitted by FJPT nodes outside the NSZ.
\subsection{Channel Models}
The wireless propagation between the HAPS and ground nodes, as well as the interference links generated by the FJPT nodes, are subject to small-scale fading and large-scale fading. For small-scale fading, 
to capture different fading conditions for aerial and terrestrial channels, we adopt the Nakagami-$m$ model with possibly different parameters for each link type \cite{shaik2024performance}. 
This model is commonly used in wireless analysis since it can represent a wide range of fading environments through the shape parameter $m$. 
In particular, measurement studies for air-to-ground channels have demonstrated that the Nakagami-$m$ distribution provides a close match to empirical channel statistics \cite{yanmaz2013achieving}.

Let $g_c$ denote the channel gain associated with the HAPS links toward ground nodes, where $c\in\{hu,he\}$ corresponds to the reference UE and  eavesdropper, respectively. 
Under the Nakagami fading assumption,  $g_c$ follows a Gamma distribution given by
% \begin{equation}
% f_{g_c}(x)=
% \frac{m_H^{m_H}}{\Gamma(m_H)\Omega_H^{m_H}}
% x^{m_H-1}\exp\!\left(-\frac{m_H}{\Omega_H}x\right), \quad x\ge0,
% \label{eq:hgain}
% \end{equation}
\begin{equation}
g_c \sim \Gamma\!\left(m_h,\frac{\Omega_h}{m_h}\right).
\label{gamm1}
\end{equation}
where $m_h$ represents the fading parameter of the aerial channel and $\Omega_h$ denotes its average power. 
Similarly, the fading associated with the terrestrial interference links from the FJPT nodes to UEs and eavesdroppers is described by the channel gains $g_v$, where $v\in\{ju,je\}$. 
These gains also follow Gamma distributions but with parameters $m_j$ and $\Omega_j$ reflecting ground propagation characteristics, i.e.,
\begin{equation}
g_v \sim \Gamma\!\left(m_j,\frac{\Omega_j}{m_j}\right).
\label{gamm2}
\end{equation}

The large-scale fading from HAPS to the reference UE/eavesdroppers is given as
\begin{equation}
\ell(d_c)=C d_c^{-\alpha_H},
\end{equation}
where $d_c, c\in\{hu,he\}$, is the distance from HAPS to the reference UE/eavesdropper, $C$ is the intercepts of the path loss formula, and  $\alpha_H$ is the path-loss exponent of the HAPS-to-ground channels.  
Considering both small-scale  and large-scale fading, the received signal power from the HAPS at a ground node $c$ is expressed as
\begin{equation}
P_{r,c} = P_h G_h g_c \ell(d_c),
\label{eq:Pr}
\end{equation}
where $P_h$ denotes the transmit power of the HAPS and $G_h$ is its antenna gain.

\subsection{Wireless Information and Power Transfer Model}
In the proposed wireless information and power transfer architecture, the receiving UE employs a time-switching mechanism to alternate between energy harvesting and information decoding \cite{ata2025time}. During each downlink transmission interval, the received RF signals are first utilized for energy harvesting over a fraction $\rho \in [0,1]$ of the time slot. The remaining portion, $(1-\rho)$, is then dedicated to information reception. Therefore,  $\rho$ represents a time allocation factor.  

During the harvesting period, the user collects RF energy from all incident transmissions in the environment. In particular, energy is harvested from the signal transmitted by the HAPS as well as from the interference generated by the FJPT nodes deployed in the network. The amount of harvested energy depends on the propagation conditions, including path loss, small-scale fading, and the directional gains resulting from the null-steering antennas used by the FJPT nodes. 
Following the energy harvesting phase, the receiver operates in the information decoding mode. 
During this stage, the HAPS transmits the confidential data signal intended for the legitimate UEs. 
At the same time, the FJPT nodes remain active and emit  interference in order to impair the reception quality at potential eavesdroppers. 
Accordingly, the typical UE attempts to decode the information transmitted by the HAPS during the remaining $(1-\rho)$ portion of the transmission slot.

\subsection{Null-Steering Antenna Model}
To mitigate the interference at the UEs while maintaining high interference at the eavesdroppers, we consider cooperation among deployed jammers. First, each FJPT node is assumed to be equipped with  null-steering antennas. Then all FJPT nodes within a distance $D$ from each UE direct nulls towards the UE while  keep radiation high interference in all other directions as illustrated in Fig.~\ref{SMfig}b. 
%Such an approach is particularly relevant in systems where the jammers are equipped with beamforming-capable antenna arrays. 
To obtain a tractable analytical model, we represent the FJPT node radiation pattern using a  sector-based model inspired by \cite{andrews2016modeling}, giving two gain levels: a suppressed gain $A_j^{\text{null}}$ inside the null region and a maximum gain $A_j^{\text{max}}$ outside the null region. 
%The null region spans an angular width $\psi$. Experimental studies have reported that practical antenna arrays can achieve null depths exceeding $20$ dB relative to the peak radiation level \cite{dicandia2016null,deng2014hemispherical}. 
For simplicity, all jammers are assumed to employ identical antenna configurations and null widths. Under this model and for a given null-width $\psi$, the antenna gain is a discrete random variable with the following model 
\begin{equation}
A_j=
\begin{cases}
A_1=A_j^{\text{null}}\;\; \text{with probability}\;\; p_1=\frac{\psi}{2\pi} & \\[6pt]
A_2=A_j^{\text{max}}\;\; \text{with probability} \;\; p_2=\frac{2\pi-\psi}{2\pi}.
\end{cases}
\label{eq:null_model}
\end{equation}

For the reference UE, the interference from FJPT nodes inside its NSZ is attenuated by $A_J^{\text{null}}$ , which, however, would reduce the amount of harvested energy during the harvesting period. 
This will result in some trade-offs with respect to the NSZ radius $D$ between the coverage performance and the harvested energy at the reference UE.
%jammers located inside its NSZ $R_0$ generate interference with the attenuated gain $A_J^{\text{null}} $. 
% In contrast, jammers outside this region may or may not place a null in the direction of the reference user, depending on the orientation of their null region toward other users. 
% As a result, the corresponding antenna gain becomes a discrete random variable
% \[
% A_J \in \{A_J^{\text{null}},A_J^{\text{tx}}\},
% \]
% with probabilities
% \[
% \mathbb{P}(A_J=A_J^{\text{null}})=\frac{\psi}{2\pi},
% \qquad
% \mathbb{P}(A_J=A_J^{\text{tx}})=\frac{2\pi-\psi}{2\pi}.
% \]
% This randomness stems from the spatially random placement of both users and jammers. 
For eavesdroppers, the interference gain of all FJPT nodes is  modelled by the same random variable $A_J$, since nulls are not intentionally directed toward eavesdroppers.

{\textit{Remark 1:}
In this work, perfect CSI and ideal beamforming are assumed, so that the nulls are accurately steered toward the legitimate UEs; the finite null gain, $A_j^{\mathrm{null}}$ already reflects the partial suppression achieved by practical null-steering. In practice, imperfect CSI and beamforming errors may shift the direction of the steered nulls, causing residual interference to leak toward the legitimate UE and thereby degrading its SINR and the resulting secrecy performance. Such imperfections can be captured within the present framework through an increased effective null depth $A_j^{\mathrm{null}}$ and direction misalignment, which is left as a direction for future work.}

{\textit{Complexity and implementation overhead.} The proposed cooperative null-steering mechanism imposes only modest overhead. Each FJPT node needs to identify the UEs within its NSZ radius $D$ and steer a spatial null toward each of them. The former is a local, low-rate operation whose signaling scales with the local UE density ($\lambda_u \pi D^2$) rather than the total network size. This keeps the coordination overhead bounded and independent of the network scale. The latter is realized through standard analog beamforming, whose complexity grows linearly with the number of antenna elements. Moreover, the scheme relies on limited channel knowledge, as the nulls are steered based only on the angular positions of the nearby UEs rather than full instantaneous channel estimates, thereby avoiding eavesdropper CSI and high-rate feedback. Being fully distributed, with each node acting only on its local neighborhood, the mechanism requires no centralized processing and its overhead remains essentially constant as the network is scaled.}
\section{Joint Rate-Energy Coverage Analysis }

In this section, we analyze the system performance in terms of the distribution of the joint rate and harvested energy at the reference UE. Let $y_m\in \Phi_j$, and $z_n\in \Phi_e$ denote the locations of the $m$-th FJPT node, and the $n$-th eavesdropper, respectively.
%\subsection{Joint Rate and Energy Coverage}
Let $\mathcal{R}_u$ and $\mathcal{E}_u$ denote the received downlink rate and the harvested energy at the reference UE, respectively. Then, the downlink rate at the reference UE, normalized by the system bandwidth, can be expressed as
\begin{equation}
\mathcal{R}_u=(1-\rho)\log_2\!\left(1+\Upsilon_u\right),
\label{rateu}
\end{equation}
where $\rho$ is the time allocation factor defined in the previous section and $\Upsilon_u$ denotes the received signal-to-interference-plus-noise ratio (SINR) at the reference UE, which is defined as
\begin{equation}
\Upsilon_u=\frac{P_{r,hu}}{I_{u}^{\mathrm{in}}+I_{u}^{\mathrm{out}}+\sigma_u^2}.
\label{sinru}
\end{equation}
{Here, $P_{r,hu}$ represents the received signal power from the transmitting HAPS at the reference UE, as defined in (\ref{eq:Pr}). The terms $I_{u}^{\mathrm{in}}$ and $I_{u}^{\mathrm{out}}$ denote the aggregate interference generated by the FJPT nodes located inside and outside the NSZ of the reference UE, respectively, while $\sigma_u^2$ is the thermal noise power. According to the proposed system model, the antenna gain towards the reference UE from all FJPT nodes inside its NSZ is $A_{j}^{\text{null}}$, while the gain of the interfering nodes outside the NSZ is a random variable given in (\ref{eq:null_model}).
As such, the interference contributions from the FJPT nodes at the reference UE can be expressed as}
\begin{equation}
I_u^{\mathrm{in}}=\sum_{y_m\in\mathcal{B}(x_o,D)}P_j A_{j}^{\text{null}} g_{ju,m}\ell(d_{ju,m}),
\label{Iin}
\end{equation}
and
\begin{equation}
I_u^{\mathrm{out}}=\sum_{y_m\in\Phi_j\setminus\mathcal{B}(x_o,D)}P_j A_j g_{ju,m}\ell(d_{ju,m}),
\label{Iout}
\end{equation}
where $\mathcal{B}(x_o,D)$ denotes a circular region centered at $x_0$ with radius $D$ and $P_j$ represents the transmit power of the FJPT nodes. The function $\ell(d_{ju,m})=C d_{ju,m}^{-\alpha_j}$ denotes the large-scale ground path-loss model, where $d_{ju,m}=||x_0-y_m||$ is the distance between the reference UE and the $m$-th FJPT node, and $\alpha_j$ is the path-loss exponent of the ground channel. 
The harvested energy, which is normalized by time, can be expressed as follows
\begin{equation}
  \mathcal{E}_u=\xi\rho (P_{r,hu}+I_u^{in}+I_u^{out}),
  \label{energyh}
\end{equation}
where $\xi$ is the efficiency of the energy converter at the UE when a linear energy harvester is considered. 
{Although most practical energy harvesters are nonlinear, we adopt the linear energy harvester model in~(\ref{energyh}) for analytical tractability as commonly adopted in system-level SWIPT studies \cite{Akin2019SWIPT,Clerckx2019JSAC}. This assumption is reasonable for the considered system because, in addition to the severe attenuation experienced by signals from distant FJPT nodes due to terrestrial path loss, the strongest nearby interfering FJPT nodes are intentionally suppressed by the proposed null-steering mechanism. Consequently, the aggregate RF power incident on the energy harvester is expected to remain within the approximately linear operating region of the harvester.}

%This is justified by the fact that the RF power received at the low-power UEs remains rather small due to the strong ground path loss, which makes the harvested power dominated by only a few nearby nodes, and the null-steering suppression of the nearest FJPT nodes, which further limits the incident power. Consequently, the received power does not reach the saturation region, and the harvesters operate within their linear range. 
%A nonlinear harvester model would be more appropriate for substantially higher jammer densities or transmit powers, which is left for future work.

For a given rate and energy threshold, $R_{th}$ and $\mathcal{E}_{th}$, the joint rate-energy coverage, which also represents the joint complementary cumulative distribution function (CCDF) of the two random variables $\mathcal{R}_u$ and $\mathcal{E}_{u}$, is defined as
\begin{equation}
F(R_{th},\mathcal{E}_{th})=\text{Pr}[\mathcal{R}_u\geq R_{th},\mathcal{E}_u\geq \mathcal{E}_{th}].
    \label{jc1}
\end{equation}

Using (\ref{rateu})-(\ref{energyh}) the joint rate-energy coverage is rewritten as
\begin{equation}
F(R_{th},\mathcal{E}_{th})=\text{Pr}\Big[\frac{\mathcal{E}_{th}}{\xi \rho}-P_{r,hu}\leq I_u^{total}\leq\frac{P_{r,hu}}{2^{\frac{R_{th}}{\rho-1}}-1}-\sigma_u^2 \Big],
    \label{jc2}
\end{equation}
where $I_u^{total}=I_u^{in}+I_u^{out}$ represent the total interference at the reference UE. 
For $P_{r,hu}>\big(\big({2^{\frac{R_{th}}{\rho-1}}-1}\big)^{-1}+1\big)(\frac{\mathcal{E}_{th}}{\xi\rho}+\sigma_u^2)$, the joint rate-energy coverage in (\ref{jc2}) is written as
\begin{eqnarray}
F(R_{th},\mathcal{E}_{th})&=&\nonumber\\
&&\!\!\!\!\!\!\!\!\!\!\!\!\!\!\!\!\!\!\!\!\!\!\!\!\!\!\!\!\!\!\!\!\!\!\!\!\!\!\!\!\!\!\!\!\!\!\!\!\!\!\!\!\!\!\!\!\!\text{Pr}\Big[I_u^{total}\leq\frac{P_{r,hu}}{2^{\frac{R_{th}}{\rho-1}}-1}-\sigma_u^2 \Big] -\text{Pr}\Big[ I_u^{total}\leq\frac{\mathcal{E}_{th}}{\xi \rho}-P_{r,hu} \Big]
    \label{jc22}
\end{eqnarray}
which can be further expressed as
\begin{eqnarray}
F(R_{th},\mathcal{E}_{th})&= \mathbb{E}_{g_{hu}} \Bigg[&F_{I_u^{total}}\bigg(\frac{P_h G_h g_{hu} \ell(d_{hu})}{2^{\frac{R_{th}}{\rho-1}}-1}-\sigma_u^2\bigg)\nonumber\\
&&\!\!\!\!\!\!\!\!\!\!\!\!\!\!\!\!\!\!\!\!\!\!\!\!\!\!\!-F_{I_u^{total}}\bigg(\frac{\mathcal{E}_{th}}{\xi \rho}-P_h G_h g_{hu} \ell(d_{hu})\bigg)\Bigg],  
  \label{jc4}
\end{eqnarray}
where $\mathbb{E}_{g_{hu}}[\cdot]$ denotes the expectation with respect to the channel gain $g_{hu}$ between the HAPS and the reference UE and $F_{I_u^{total}}(x)=\text{Pr}[I_u^{total}\leq x]$ is the CDF of $I_u^{total}$ at threshold $x$. Since $I_{u}^{in}$ and $I_{u}^{out}$ are independent for a given NSZ radius $D$, the interference CDFs in (\ref{jc4}) can be computed using  the Gil-Pelaez inversion
theorem \cite{di2014stochastic} as follows
\begin{eqnarray}
F_{I_{u}^{total}}(x)&=& F_{I_{u}^{in},I_{u}^{out}}(x)=\frac{1}{2}\nonumber\\
&&\!\!\!\!\!\!\!\!\!\!\!\!\!\!\!\!\!\!\!\!\!\!\!\!\!\!\!\!\!\!\!\!\!\!\!\!-\frac{1}{\pi}\int_{0}^{\infty}
\frac{\mathrm{Im}\!\left\{\exp(-j\omega x)\Theta_{I_{u}^{in}}(j\omega)\Theta_{I_{u}^{out}}(j\omega)\right\}}{\omega}\,d\omega,
\label{gil}
\end{eqnarray}
where  $\mathrm{Im}\{\cdot\}$ denotes the imaginary part of a complex quantity, $j=\sqrt{-1}$, and $\Theta_{I_{u}^{in}}(j\omega)$ and $\Theta_{I_{u}^{out}}(j\omega)$, respectively, represent the characteristic functions of the interference $I_u^{in}$ and $I_u^{out}$ received at the reference UE. From (\ref{jc4}) and (\ref{gil}), it is clear that obtaining a closed-form expression for the joint rate-energy coverage  requires the characterization of the interference through its characteristic functions and averaging the resulting expression over the small-scale fading channel gain $g_{hu}$. The distribution of $g_{hu}$ is characterized by the following probability density function (PDF)
\begin{eqnarray}    
f_{g_{hu}}(x)=
\frac{m_h^{m_h}}{\Gamma(m_h)\Omega_h^{m_h}}
x^{m_h-1}
\exp\!\left(-\frac{m_h x}{\Omega_h}\right), \quad x\ge0
\label{gPDF}
\end{eqnarray}

The characteristic function can be derived for $I_u^{in}$ and $I_u^{out}$ according to the following lemma.

\textit{Lemma 1.} For a given NSZ radius $D$, the characteristic functions of the interference $I_u^{in}$ and $I_u^{out}$ at the reference UE, denoted as   $\Theta_{I_{u}^{\mathrm{in}}}(j\omega)$ and $\Theta_{I_{u}^{\mathrm{out}}}(j\omega)$, are respectively given in (\ref{CF1}) and (\ref{CF2}) at the top of the next page, where $G^{m,n}_{p,q}(\cdot)$ denotes the Meijer's $G$-function \cite[Eq.~(9.301)]{gradshteyn2014table}.

\begin{figure*} [!t]
\begin{eqnarray}
\Theta_{I_u^{in}}(jw) & =&
     \exp\Bigg( -
     \pi \lambda_j  
     D^2 \Bigg(1-\frac{\Gamma\!\left(1-\frac{2}{\alpha_j}\right)}
{\Gamma(m_j)\Gamma\!\left(-\frac{2}{\alpha_j}\right)}
\,G^{\,1,2}_{2,2}\!\left(
\begin{matrix}
1-m_j,\;1+\frac{2}{\alpha_j}\\
0,\;\frac{2}{\alpha_j}
\end{matrix}
\bigg|\, 
     \frac{-jw CP_j A_j^{\text{null}}D^{-\alpha_j}\Omega_J}{m_j }
\right)\Bigg)\Bigg),
\label{CF1}
\end{eqnarray}
\end{figure*} 
\begin{figure*} [!t]
\begin{eqnarray}
  \Theta_{I_u^{out}}(jw) & =&
     \exp\Bigg(-
            \pi \lambda_J \sum_{q=1}^2 p_{q} \Bigg\{ R_s^2
     \Bigg(1-\frac{\Gamma\!\left(1-\frac{2}{\alpha_j}\right)}
{\Gamma(m_j)\Gamma\!\left(-\frac{2}{\alpha_j}\right)}
\,G^{\,1,2}_{2,2}\!\left(
\begin{matrix}
1-m_j,\;1+\frac{2}{\alpha_j}\\
0,\;\frac{2}{\alpha_j}
\end{matrix}
\bigg|\,  
     \frac{-jw CP_j A_q R_s^{-\alpha_j}\Omega_J}{m_j }
\right)\Bigg) \nonumber\\
     &&
     -D^2 \Bigg(1-\frac{\Gamma\!\left(1-\frac{2}{\alpha_j}\right)}
{\Gamma(m_j)\Gamma\!\left(-\frac{2}{\alpha_j}\right)}
\,G^{\,1,2}_{2,2}\!\left(
\begin{matrix}
1-m_j,\;1+\frac{2}{\alpha_j}\\
0,\;\frac{2}{\alpha_j}
\end{matrix}
\bigg|\,  
     \frac{-jw CP_j A_qD^{-\alpha_j}\Omega_J}{m_j }
\right)\Bigg) 
     \Bigg\}\Bigg),
     \label{CF2}
     \end{eqnarray}
%     \hline
\end{figure*}
\textit{Proof: See Appendix A}

The joint rate-energy coverage  at the reference UE can now be formally characterized, as stated in the following proposition.

\textit{Proposition 1.} For a given NSZ radius, $D$, the joint rate-energy coverage is expressed as in (\ref{JC6}) 
\begin{figure*} [!t]
\begin{eqnarray}
F(R_{th},\mathcal{E}_{th}) & =&
\frac{1}{\pi}\int_0^\infty \text{Im}\Bigg\{\frac{1}{\omega}\Bigg(e^{-j\omega \frac{\mathcal{E}_{th}}{\xi \rho}}
\left(1-\frac{j\omega P_h G_h \ell(a) \Omega_h}{m_h}\right)^{-m_h}-
e^{j\omega \sigma_u^{2}}
\left(1+\frac{j\omega P_h G_h \ell(a) \Omega_h}{m_h (2^{\frac{R_{th}}{\rho-1}}-1)}\right)^{-m_h}\Bigg)\nonumber\\
&&
    \exp\Bigg( -
     \pi \lambda_j  
     D^2 \mathcal{M}\Bigg( 
     \frac{jw CP_j A_j^{\text{null}}D^{-\alpha_j}\Omega_J}{m_j }
\Bigg)\Bigg) \exp\Bigg(-
            \pi \lambda_J \sum_{q=1}^2 p_{q} \Bigg\{ R_s^2
     \mathcal{M}\Bigg( 
     \frac{jw CP_j A_q R_s^{-\alpha_j}\Omega_J}{m_j }
\Bigg) \nonumber\\
&&
     -D^2 \mathcal{M}\Bigg(  
     \frac{jw CP_j A_qD^{-\alpha_j}\Omega_J}{m_j }
\Bigg) 
     \Bigg\}\Bigg)\Bigg\}d\omega,
\label{JC6}
\end{eqnarray}
%\hline
\end{figure*}
at the top of the next page, where $\mathcal{M}(x)$ is given as
\begin{eqnarray}
  \mathcal{M}(x)=  1-\frac{\Gamma\!\left(1-\frac{2}{\alpha_j}\right)}
{\Gamma(m_j)\Gamma\!\left(-\frac{2}{\alpha_j}\right)}
\,G^{\,1,2}_{2,2}\!\left(
\begin{matrix}
1-m_j,\;1+\frac{2}{\alpha_j}\\
0,\;\frac{2}{\alpha_j}
\end{matrix}
\bigg|\, 
     x
\right).
\label{Mfun}
\end{eqnarray}

\textit{Proof:} Applying (\ref{gil}) in (\ref{jc4}) and averaging over the channel gain $g_{hu}$ using its PDF in (\ref{gPDF}), we get the following expression
\begin{eqnarray}
F(R_{th},\mathcal{E}_{th})&=& \frac{1}{\pi}\int_0^{\infty}\int_{0}^{\infty}
\frac{1}{\omega}\mathrm{Im}\Bigg\{ \Theta_{I_{u}^{in}}(j\omega)\Theta_{I_{u}^{out}}(j\omega)\Bigg(\nonumber\\
&&\!\!\!\!\!\!\!\!\!\!\!\!\!\!\!\!\!\!\!\!\!\!\!\!\!\!\!\!\!\!\!\!\!\!\!\!\exp\bigg(-j\omega \bigg(\frac{\mathcal{E}_{th}}{\xi \rho}-P_h G_h g_{hu} \ell(d_{hu})\bigg)\bigg) \nonumber\\
&&\!\!\!\!\!\!\!\!\!\!\!\!\!\!\!\!\!\!\!\!\!\!\!\!\!\!\!\!\!\!\!\!\!\!\!\!\!\!\!\!\!\!\!\!\!\!\!\!\!\! -\exp\bigg(\!\!\!-j\omega \bigg(\frac{P_h G_h g_{hu} \ell(d_{hu})}{2^{\frac{R_{th}}{\rho-1}}-1}-\sigma_u^2\!\!\bigg)\!\!\bigg)\!\!\Bigg)\!\!\Bigg\}f_{g_{hu}}(g_{hu})d\omega dg_{hu}.
\label{proof_jc}
\end{eqnarray}
Then, by interchanging the order of integration and integrating with respect to $g_{hu}$, followed by substituting $\mathcal{M}(x)$ as in (\ref{Mfun}), we obtain the expression in (\ref{JC6}). 

Since (\ref{JC6}) represents the joint CCDF of $\mathcal{R}_u$ and $\mathcal{E}_u$, we can obtain the individual rate and  energy coverage expressions as marginal CCDFs such that  $F_{\mathcal{R}_u}(R_{th})=F(R_{th},\mathcal{E}_{th})$ for $\mathcal{E}_{th}=0$ and 
$F_{\mathcal{E}_u}(\mathcal{E}_{th})=F(R_{th},\mathcal{E}_{th})$ for $R_{th}=0$.

{We note that owing to the near-vertical stratospheric geometry within a single SVR, the elevation angles are large and the LoS probability is high across this region. Using the LOS probability model in~\cite{al2020modeling}, it exceeds $97\%$ even at $2$~km from the SVR center for urban areas. The NLoS probability is therefore negligible and is ignored, so the HAPS-to-ground links are accurately captured by the strong-LoS model. The framework however can be extended to an explicit LoS/NLoS model by averaging the derived coverage expressions over the two states, with the ground-based jamming coordination ($D$, null-steering, and jammer density) remaining the primary driver of the performance trade-offs.}

{
\section{Secrecy Rate Analysis}
We characterize the secrecy performance   of the considered system in terms of the average secrecy rate, which is defined as the difference between the achievable rate at the legitimate UE and that at the most detrimental eavesdropper, i.e., the one with the maximum SINR, \cite{huang2015secure},
% It is defined as
\begin{equation}
\overline{\mathcal{R}}_s
=
\mathbb{E}\!\left[\max\Big(\mathcal{R}_{u}-\mathcal{R}_{e},0\Big)\right],
\label{secCap1}
\end{equation}
where $\mathcal{R}_u$ denote the rate at the reference UE which is given in (\ref{rateu}) and $\mathcal{R}_e$ is the received rate at the most detrimental eavesdropper. Since confidential information is transmitted only during the information decoding phase, $\mathcal{R}_e$ is defined  as 
\begin{equation}
\mathcal{R}_e=(1-\rho)\log_2\!\left(1+\Upsilon_{e,M}\right),
\label{rate3e}
\end{equation}
with
\begin{equation}
\Upsilon_{e,M}=\max_{k\in\Phi_e}\Upsilon_{e,k},
\end{equation}
where $\Upsilon_e$ denotes the received SINR at a given eavesdropper, which is expressed as
\begin{equation}
\Upsilon_e=\frac{P_{r,he}}{I_{e}+\sigma_e^2},
\label{sinre}
\end{equation}
where $P_{r,he}$ denotes  the signal power received  by the  eavesdropper from the transmitting HAPS, as defined in (\ref{eq:Pr}),  $I_{e}$ represents the aggregate interference generated by  all FJPT nodes, and  $\sigma_e^2$ is the thermal noise power at the eavesdropper. The interference $I_{e}$ can be expressed as
\begin{equation}
I_e=\sum_{y_m\in\Phi_j}P_j A_j g_{je,m}\ell(d_{je,m}),
\end{equation}
where $d_{je,m}=||z_0-y_m||$ is the distance between the  eavesdropper and the $m$-th FJPT node and $g_{je,m}$ is the channel gain its distribution is defined in (\ref{gamm2}).
The expectation in \eqref{secCap1} can be written as
\begin{equation}
\overline{R}_s
=\mathbb{E}_{\Upsilon_u,\Upsilon_{e,M}}\bigg[\max\bigg(\log_2\bigg(\frac{1+\Upsilon_u}{1+\Upsilon_{e,M}}\bigg),0\bigg)\bigg].
\end{equation}
\\
To proceed, we need first to find the cumulative distribution functions (CDFs) of $\Upsilon_u$ and $\Upsilon_{e,M}$ denoted as $F_{\Upsilon_u} (\gamma_u)$ and $F_{\Upsilon_{e,M}}(\gamma_e)$, respectively. According to \cite{andrews2016modeling}, for an integer $m_h$, the approximated closed form of $F_{\Upsilon_u}(\gamma_u)$  can be obtained as
\begin{eqnarray}
% \begin{aligned}
 F_{\Upsilon_u}(\gamma_u)&\!\!\!\!\!\!=\!\!\!\!\!\!&1-\sum_{b=1}^{m_h}(-1)^{b+1}\binom{m_h}{b} \mathcal{T}_{I_u^{in}}\!\!\left(\frac{b \kappa \gamma_u}{P_h G_h  C \ell(a)}\right) \nonumber\\
&&\!\!\!\!\!\!\!\!\!\!\!\!\!\!\!\!\!\!\!\!\!\!\!\!\times\mathcal{T}_{I_u^{out}}\!\!\left(\frac{b \kappa \gamma_u}{P_h G_h  C \ell(a)}\right)\exp\left(\frac{v \kappa \gamma_u\sigma_u^2}{P_h G_h  C \ell(a)} \right),
\label{cdfu}
\end{eqnarray} 
where $\kappa=\frac{m_h}{\Omega_h} \Gamma(m_h+1)^{\frac{-1}{m_h}}$, $\mathcal{T}_{I_u^{in}}(s)$ and $\mathcal{T}_{I_u^{out}}(s)$ are the Laplace transforms of the interference at the reference UE, $I_u^{in}$ and $I_u^{out}$, respectively. Since the interference is a non-negative random variable, the Laplace transforms of $I_u^{in}$ and $I_u^{out}$ can be obtained from their characteristic functions given in (\ref{CF1}) and (\ref{CF2}) such that $\mathcal{T}_{I_u^{in}}(s)=\Theta_{I_u^{in}}(jw)_{jw\to -s}$ and $\mathcal{T}_{I_u^{out}}(s)=\Theta_{I_u^{out}}(jw)_{jw\to -s}$.
The CDF of $\Upsilon_{e,M}$
 is given as
\begin{eqnarray}
F_{\Upsilon_{e,M}}(\gamma_e)=\text{Pr}\big[\Upsilon_{e,M}\leq \gamma_e\big]=\text{Pr}\Big[\max_{k\in \Phi_E} \Upsilon_{e,k}\leq \gamma_e\Big].
\end{eqnarray}
 By applying the probability generating functional
(PGFL) \cite{haenggi2012stochastic}, we get  
\begin{eqnarray}
F_{\Upsilon_{e,M}}(\gamma_e)&=&\mathbb{E}\Big(\prod_{k\in \Phi_e} \text{Pr}\left[\Upsilon_{e,k}\leq \gamma_e\right]\Big)\nonumber\\
&&\!\!\!\!\!\!\!\!\!\!\!\!\!\!\!\!\!\!\!\!=\exp\Big(-\lambda_e\int_{\mathbb{R}^2}\left(1-\text{Pr}\left[\Upsilon_{e,k}\leq \gamma_e\right]\right)dr_k\Big)
\end{eqnarray}
Using polar coordinates over the SVR gives
\begin{eqnarray}
F_{\Upsilon_{e,M}}(\gamma_e)=\exp\Big(-2\pi \lambda_e\int_{0}^{R_s}\left(1-F_{\Upsilon_{e}|r}(\gamma_e)\Big)rdr\right)
\label{maxSINR_sp}
\end{eqnarray}
where $F_{\Upsilon_{e}|r}(\gamma_e)$ denotes the CDF of $\Upsilon_e$ for an eavesdropper located at a distance $r$ from the reference UE. Similarly, 
according to \cite{andrews2016modeling}, for an integer $m_h$, $F_{\Upsilon_{e}|r}(\gamma_e)$  can be obtained as
\begin{eqnarray}
% \begin{aligned}
 F_{\Upsilon_e|r}(\gamma_e)\!\!\!\!\!\!&=\!\!\!\!\!\!&1-\sum_{b=1}^{m_h}(-1)^{b+1}\binom{m_h}{b}\times\nonumber\\ &&\!\!\!\!\!\!\!\!\!\!\!\!\!\!\!\!\!\!\!\!\!\!\!\!\!\!\!\!\!\!\!\!\!\!\!\!\!\!\!\!\!\! \mathcal{T}_{I_e|r}\!\!\left(\frac{b \kappa \gamma_e}{P_h G_h  C \ell(\sqrt{a^2+r^2})}\right)\! \exp\!\!\left(\frac{v \kappa \gamma_e\sigma_e^2}{P_h G_h  C \ell(\sqrt{a^2+r^2})} \right),
\label{Pcov2}
\end{eqnarray} 
where $\mathcal{T}_{I_e|r}(s)$ is the Laplace transform of the interference at the  eavesdropper, which can be obtained as in the following lemma.
\\
\textit{Lemma 2.} For a given NSZ radius $D$, the Laplace transform of the interference at an  eavesdropper located at a distance $r$ from the reference UE, $\mathcal{T}_{I_e|r}(s)$ is given in (\ref{LT3})  at the top of the next page.  
\begin{figure*} [!t]
\begin{eqnarray}
  \mathcal{T}_{I_e|r}(s) \!\!\!\!\!\!& =\!\!\!\!\!&
     \exp\Bigg(\!\!\!-
            \pi \lambda_J \sum_{q=1}^2 p_{q} \Bigg\{\!\! (R_s-r)^2
     \Bigg(\!\!1-\frac{\Gamma\!\left(1-\frac{2}{\alpha_j}\right)}
{\Gamma(m_j)\Gamma\!\left(-\frac{2}{\alpha_j}\right)}
\,G^{\,1,2}_{2,2}\!\left(
\begin{matrix}
1-m_j,\;1+\frac{2}{\alpha_j}\\
0,\;\frac{2}{\alpha_j}
\end{matrix}
\bigg|\,  
     \frac{-s CP_j A_q (R_s-r)^{-\alpha_j}\Omega_J}{m_j }
\!\!\right)\!\!\Bigg)\!\!\Bigg\}\!\Bigg).
     \label{LT3}
     \end{eqnarray}
  %   \hline
\end{figure*}
\\
\textit{Proof:} The expression in (\ref{LT3}) can be derived by following steps similar to those used in the proofs of (\ref{CF1}) and (\ref{CF2})  in Lemma 1. However, it is necessary to account for the fact that the antennas of all FJPT nodes may randomly produce either a null or the maximum gain in the direction of the reference eavesdropper. Moreover, the  eavesdropper is assumed to be located at a distance $r$ from the center of the SVR.}

\begin{figure}[t]
    \centering
\includegraphics[scale=0.65]{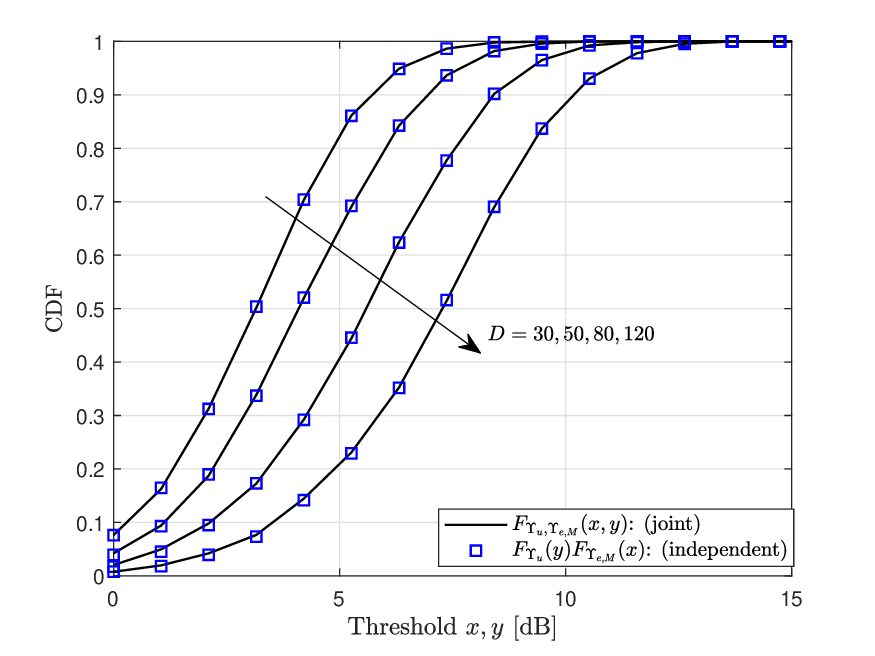}
    \caption{{Joint CDF of the reference UE and eavesdropper SINRs, $F_{\Upsilon_u,\Upsilon_{e,M}}(x,y)$, versus the product of the marginal CDFs, $F_{\Upsilon_u}(y)F_{\Upsilon_{e,M}}(x)$, for different values of the NSZ radius $D$ and $\rho=0.2$. All other system parameters are given in Section V.}}
    \label{CDFsComp}
    \end{figure}
 {It is worth mentioning that in the proposed  setup,  the small-scale fading of the aerial and ground links are all independent. Furthermore, under the proposed
 null-steering and coordination scheme, the strongest interferers within the
NSZ of each UE are suppressed. This makes the  set of dominant interferers observed by the reference UE totally different from  
 those observed by the most detrimental eavesdropper.
Consequently, the correlation between  $\Upsilon_e$ and $\Upsilon_{e,M}$ is negligible and can be ignored. Fig.~\ref{CDFsComp} compares the joint CDF $F_{\Upsilon_u,\Upsilon_{e,M}}(x,y)$, obtained by simulating both SINRs from the same HAPS transmission and the same realization of the jammer field, against the product of the marginal CDFs $F_{\Upsilon_u}(x)F_{\Upsilon_{e,M}}(y)$, for several values of the NSZ radius $D$. The two curves are virtually indistinguishable over the entire range and for all values of $D$, which confirms that $\Upsilon_u$ and $\Upsilon_{e,M}$ are effectively independent.}

{Therefore, similar to the  procedure as in \cite{wang2014physical}, the  average secrecy rate for a given distance $r$ between the reference UE and eavesdropper is   expressed as
\begin{equation}
\overline{R}_{s}
=
\frac{1}{\ln 2}
\int_{0}^{\infty}
\frac{F_{\Upsilon_u}(x)
 F_{\Upsilon_{e,M}}(x)}
{1+x}\,dx,
\label{secCap2}
\end{equation}
%where $\varpi \in \{\mathrm{NJD},\mathrm{RJD},\mathrm{NSJ}\}$ indicates the adopted coordination strategy.}
where $F_{\Upsilon_u}(x)$ and 
 $F_{\Upsilon_{e,M}}(x)$ are given in (\ref{cdfu}) and  (\ref{maxSINR_sp}), respectively.}%  can be computed as follows.
\section{Numerical Results}

In this section, we numerically evaluate the performance of the proposed secure wireless data reception and energy harvesting system. The evaluation considers several key performance metrics, including the individual and joint rate-energy coverage probabilities at a legitimate wirelessly powered UE, as well as the secrecy performance in the presence of spatially distributed eavesdroppers. The results aim to highlight how FJPT nodes can cooperatively operate in HAPS-based systems to simultaneously provide secure communication and wireless power transfer under practical deployment conditions.

Unless otherwise stated, the system parameters used in this section are set as follows. A single HAPS is deployed at an altitude of $a = 20~\text{km}$ and provides coverage over a wide ground area through multi-beam transmission. We consider a single SVR located directly beneath the HAPS with a radius of $R_s = 1~\text{km}$, which can be covered by a single beam \cite{3gpp20183rd}. The HAPS transmits with power $P_h = 50~\text{dBm}$ and antenna gain $G_h = 40~\text{dB}$. Ground users are distributed according to a homogeneous PPP with density $\lambda_u = 10^{-4}~\text{users/m}^2$, while eavesdroppers and FJPT nodes follow independent PPPs with densities $\lambda_e = 5 \times 10^{-5}~\text{eavesdroppers/m}^2$ and $\lambda_j = 10^{-3}~\text{jammers/m}^2$, respectively. Each FJPT node transmits with power $P_J = 5~\text{dBm}$ and is equipped with a directional antenna characterized by a maximum gain $A_j^{\text{max}} = 10~\text{dB}$, a null gain $A_j^{\text{null}} = -20~\text{dB}$, and a null-width of $\psi = 30^\circ$. The HAPS-to-ground channel is modeled using Nakagami-$m$ fading with parameters $(\alpha_h, m_h, \Omega_h) = (2.2, 3, 1/3)$, while the ground-based jamming links are characterized by $(\alpha_j, m_j, \Omega_j) = (2.2, 2, 1/2)$. 
The system operates at a carrier frequency of $2~\text{GHz}$ with a bandwidth of $20~\text{MHz}$. The normalized rate and energy thresholds are set to ${R}_{\mathrm{th}} = 1~\text{bps/Hz}$ and $\mathcal{E}_{\mathrm{th}} = -40~\text{dBm}$, respectively.

Performance results are obtained via Monte Carlo simulations. The simulated results are compared with the analytical expressions derived in Section III to validate their accuracy and to demonstrate the effectiveness of the proposed system model, while also providing useful insights into system behavior.
\begin{figure}[t]
    \centering
\includegraphics[scale=0.65]{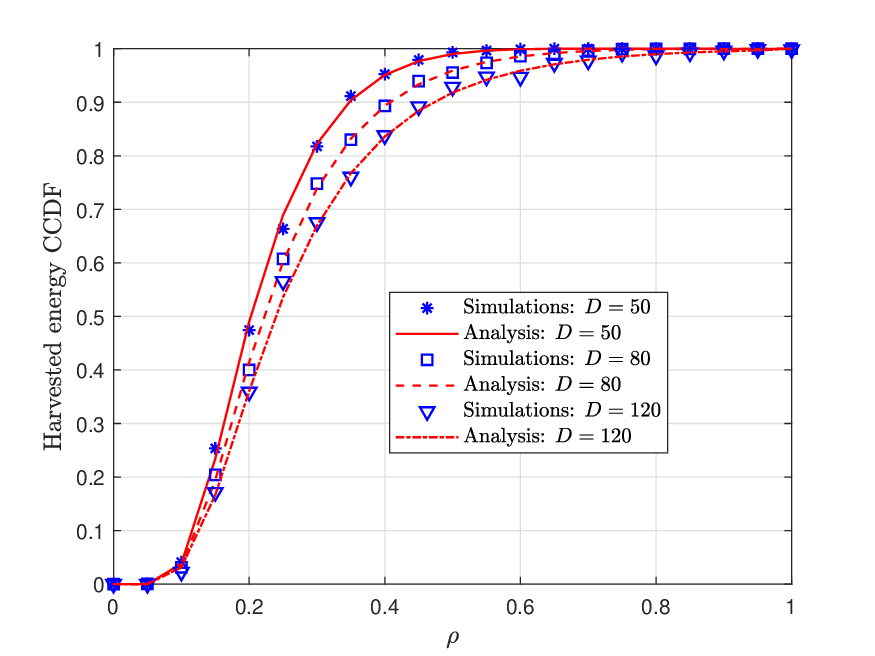}
    \caption{Harvested energy coverage versus the time allocation factor, $\rho$, for different values of the NSZ radius.}
    \label{EvsSp}
    \end{figure}

    \begin{figure}[t]
    \centering
\includegraphics[scale=0.65]{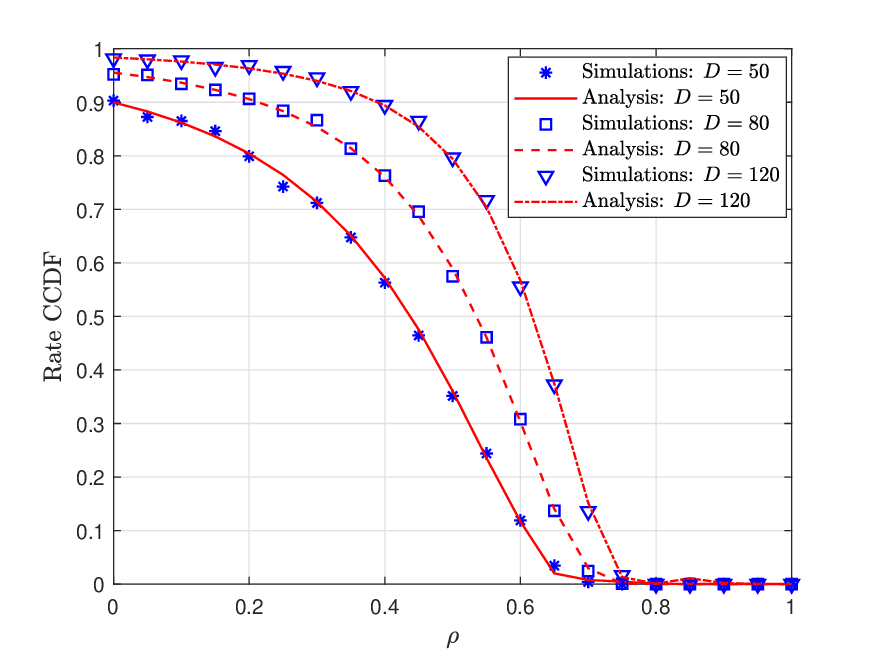}
    \caption{Rate coverage versus the time allocation factor, $\rho$, for different values of the NSZ radius.}
    \label{RvsSp}
    \end{figure}

In Fig.~\ref{EvsSp}, we illustrate the energy coverage probability as a function of the time allocation factor $\rho$ for three values of the NSZ radius, $D = 50$, $80$, and $120$ m. The results are obtained for an energy threshold of $\mathcal{E}_{\mathrm{th}} = -40~\text{dBm}$. As observed, the energy coverage improves monotonically with $\rho$, since allocating a larger fraction of time to energy harvesting increases the amount of collected RF energy. 
On the other hand, the energy coverage $F_{\mathcal{E}_u}(\mathcal{E}_{\mathrm{th}})$ decreases as $D$ increases. This behavior is attributed to the expansion of the NSZ around the user, which suppresses nearby high-power interfering FJPT nodes. While this interference mitigation is beneficial for data reception, it simultaneously reduces the available RF energy for harvesting within a given time slot. 
Furthermore, Fig.~\ref{EvsSp} demonstrates an accurate match between simulation and analytical results, confirming the accuracy of the derived expression for the energy coverage, $F_{\mathcal{E}_u}(\mathcal{E}_{\mathrm{th}})$, which is the marginal CCDF obtained from~(\ref{JC6}) by setting $R_{\mathrm{th}} = 0$.

Fig.~\ref{RvsSp} presents the rate coverage as a function of $\rho$ for the same set of NSZ radii and a rate threshold of $R_{\mathrm{th}} = 1~\text{bps/Hz}$. In contrast to the energy coverage behavior, the rate coverage decreases with increasing $\rho$ and improves with larger values of $D$. This behavior can be explained as follows: increasing $\rho$ reduces the fraction of time allocated for data transmission, thereby lowering the achievable rate. Meanwhile, enlarging the NSZ suppresses dominant nearby interferers, leading to an improvement in the received SINR and, consequently, the achievable rate at the reference UE. 
Similar to the previous case, Fig.~\ref{RvsSp} shows a close match between simulation and analytical results, validating the accuracy of the rate coverage expression $F_{\mathcal{R}_u}(R_{\mathrm{th}})$, which is obtained as a marginal CCDF from~(\ref{JC6}) by setting $\mathcal{E}_{\mathrm{th}} = 0$.

\begin{figure}[t]
    \centering
\includegraphics[scale=0.65]{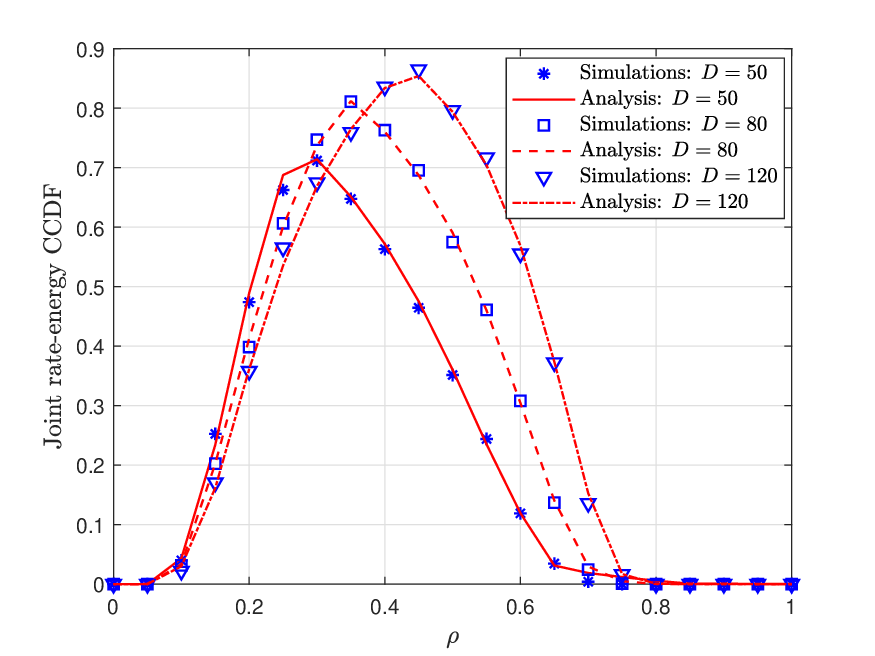}
    \caption{Joint rate-energy coverage versus the time allocation factor, $\rho$, for different values of the NSZ radius.}
    \label{JREvsSp}
    \end{figure}

The results in Fig. \ref{JREvsSp} shows the joint rate--energy coverage as a function of the time allocation factor $\rho$ for different values of the NSZ radius $D$ at rate threshold $R_{th}=1$ bps/Hz and energy threshold $\mathcal{E}_{th}=-40$ dBm. The curves exhibit a non-monotonic behavior. For small $\rho$, the joint coverage is low due to insufficient energy harvesting. As $\rho$ increases, the harvested energy improves, leading to higher coverage. However, for large $\rho$, the performance degrades since less time is allocated for data transmission. 
In addition, increasing $D$ improves the performance. This is because a larger NSZ suppresses strong nearby interference from friendly jammers, which enhances the SINR during the decoding phase while still allowing sufficient energy to be harvested. It is clear that for a given $D$, there is an optimal value of $\rho$, which provide maximum joint rate-energy coverage, and this optimal value increases with $D$. 
Finally, the analytical and simulation results closely match, validating the accuracy of the derived expression of $F(R_{th},\mathcal{E}_{th})$ provided in Proposition 1.

\begin{figure}[t]
    \centering
\includegraphics[scale=0.65]{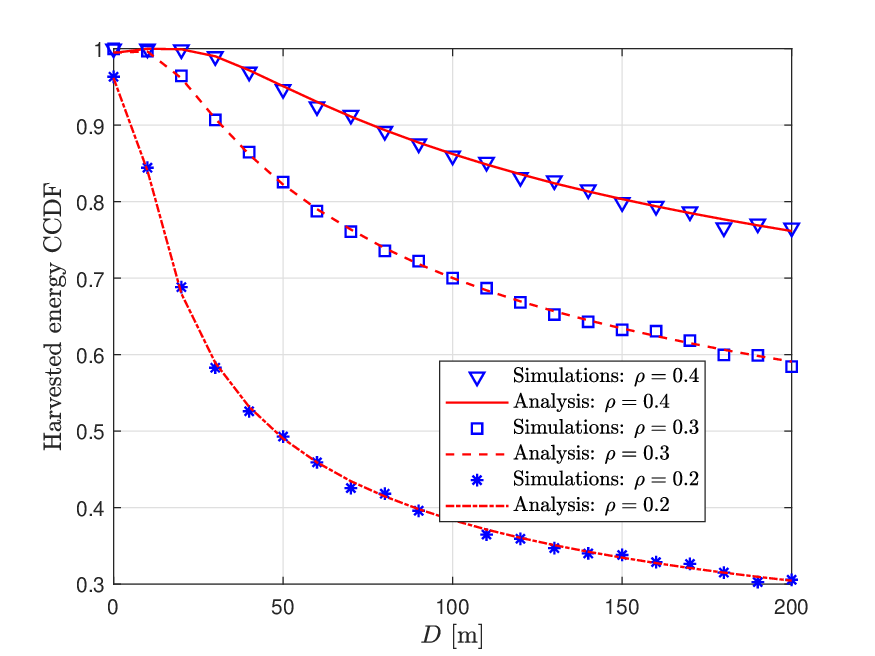}
    \caption{Harvested energy coverage versus the NSZ radius, $D$, for different values of the time allocation factor, $\rho$.}
    \label{EvsD}
    \end{figure}

    \begin{figure}[t]
    \centering
\includegraphics[scale=0.65]{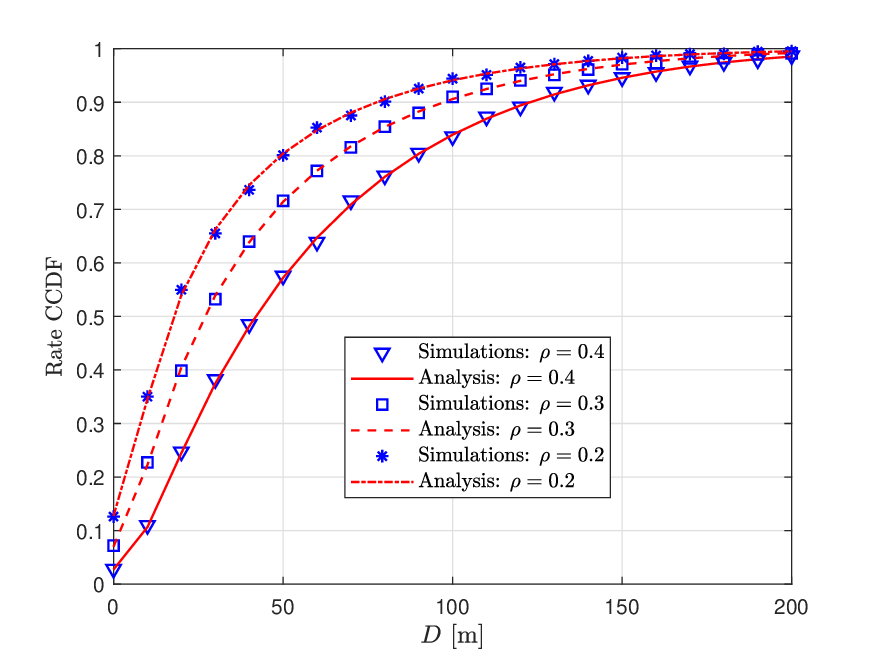}
    \caption{Rate coverage versus the NSZ radius, $D$, for different values of the time allocation factor, $\rho$.}
    \label{RvsD}
    \end{figure}

Figs.~\ref{EvsD} and \ref{RvsD} illustrate the harvested energy coverage and rate coverage, respectively, as functions of the NSZ radius $D$ for different values of the time allocation factor $\rho$. 
In Fig.~\ref{EvsD}, the harvested energy coverage monotonically decreases with $D$. This is because increasing the NSZ radius enlarges the region where nearby friendly jammers suppress their transmissions toward the user, thereby reducing the aggregate received RF power available for energy harvesting. Moreover, higher values of $\rho$ yield better energy coverage since a larger fraction of time is allocated to energy harvesting. 
On the other hand, Fig.~\ref{RvsD} shows that the rate coverage improves with increasing $D$. This is due to the reduction of strong nearby interference caused by null-steering, which enhances the SINR during the information decoding phase. Additionally, smaller values of $\rho$ provide higher rate coverage, as more time is dedicated to data transmission.

% These results highlight a fundamental trade-off: increasing $D$ benefits the rate performance but degrades energy harvesting, while increasing $\rho$ enhances energy coverage at the expense of rate performance. Finally, the analytical results closely match the simulations in both figures, confirming the accuracy of the proposed framework.

\begin{figure}[t]
    \centering
\includegraphics[scale=0.65]{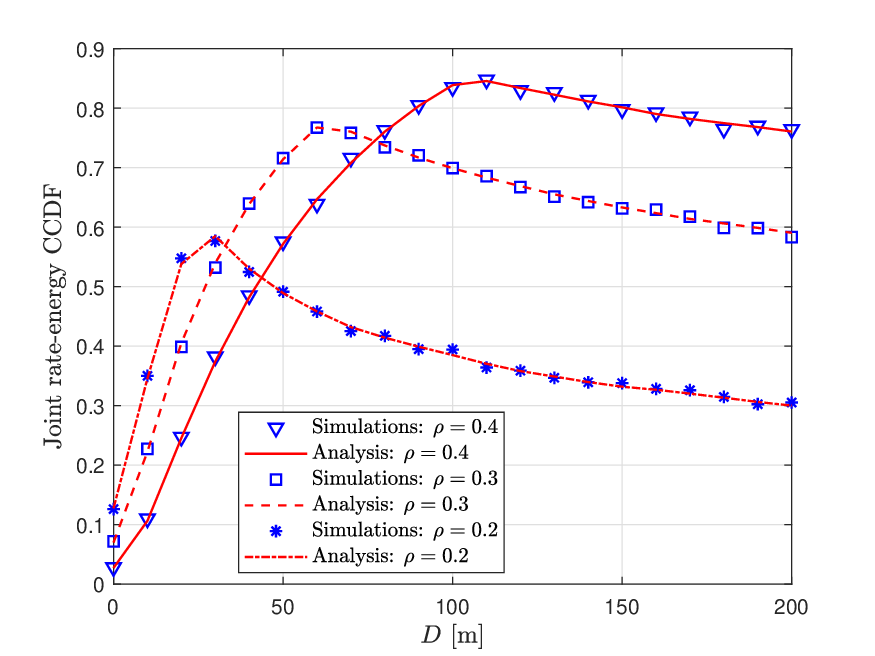}
    \caption{Joint rate and energy coverage versus the NSZ radius, $D$, for different values of the time allocation factor, $\rho$.}
    \label{JREvsD}
    \end{figure}

Fig.~\ref{JREvsD} shows the joint rate-energy coverage versus the NSZ radius $D$ for different values of $\rho$ at  rate and energy thresholds $R_{th}=1$ bps/Hz and  $\mathcal{E}_{th}=-40$ dBm. 
The performance also exhibits a non-monotonic trend with $D$. These results highlight a fundamental trade-off: increasing $D$ benefits the rate performance but degrades energy harvesting, while increasing $\rho$ enhances energy coverage at the expense of rate performance. As shown in the figure, for a fixed $\rho$, the joint rate-energy coverage is maximized at a specific value of $D$, and this optimal $D$ shifts to larger values as $\rho$ increases.

    \begin{figure}[t]
    \centering
\includegraphics[scale=0.65]{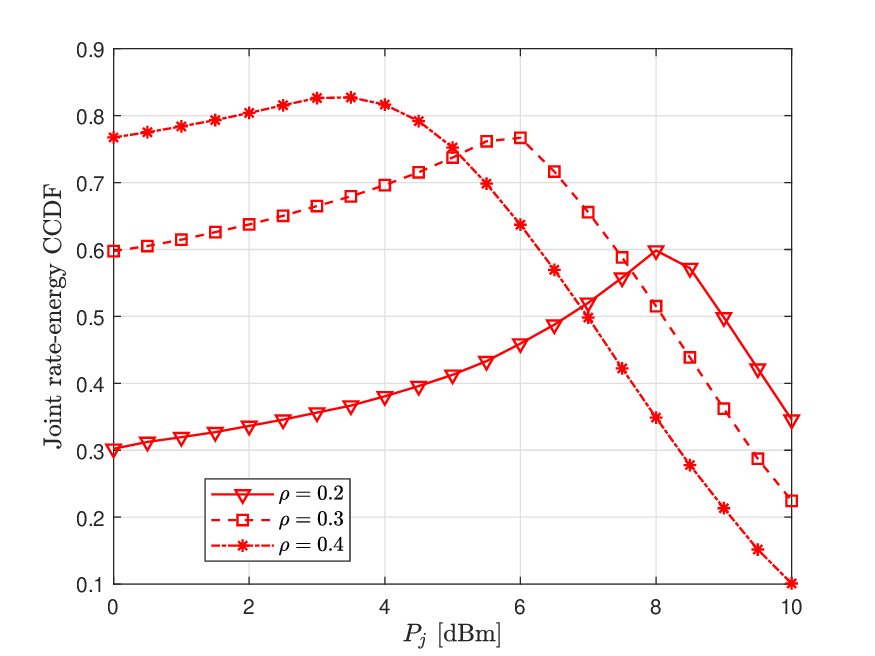}
    \caption{Joint rate and energy coverage versus the transmit power of the deployed FJPT nodes, $P_j$, for different values of the time allocation factor, $\rho$, at a NSZ radius $D=80$ m.}
    \label{JREvsPj}
    \end{figure}

Fig.~\ref{JREvsPj} shows the joint rate-energy coverage as a function of the FJPT nodes transmit power $P_j$ for different values of $\rho$. 
The performance improves with $P_j$ at low power levels due to increased harvested energy at the user. However, beyond a certain point, further increasing $P_j$ leads to a degradation in performance as the interference becomes dominant and significantly reduces the SINR and hence the rate during the decoding phase. 
In addition, the optimal $P_j$ shifts to higher values as $\rho$ decreases. This is because higher transmit power is required to compensate for the reduced harvesting duration.

%Finally, the results highlight the trade-off between energy harvesting and interference, and the analytical curves closely match the simulations, confirming the accuracy of the proposed model.

\begin{figure}[t]
    \centering
\includegraphics[scale=0.65]{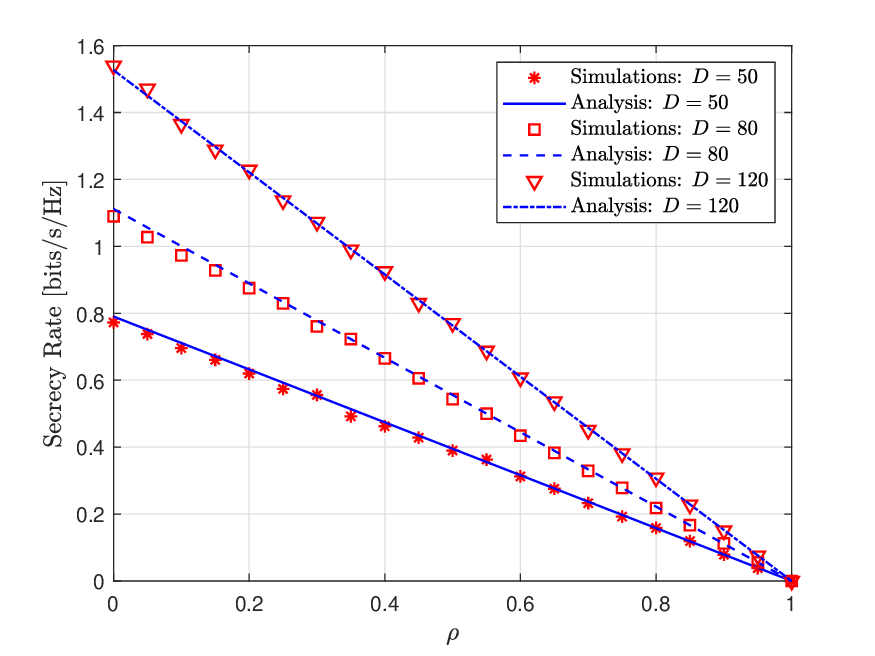}
    \caption{{Secrecy rate versus the time allocation factor, $\rho$, for different values of  the NSZ radius, $D$. %In this figure, we consider the most detrimental eavesdropper, i.e., the one with the maximum SINR.
    } }
    \label{SRvsSp}
    \end{figure}
    
\begin{figure}[t]
    \centering
\includegraphics[scale=0.65]{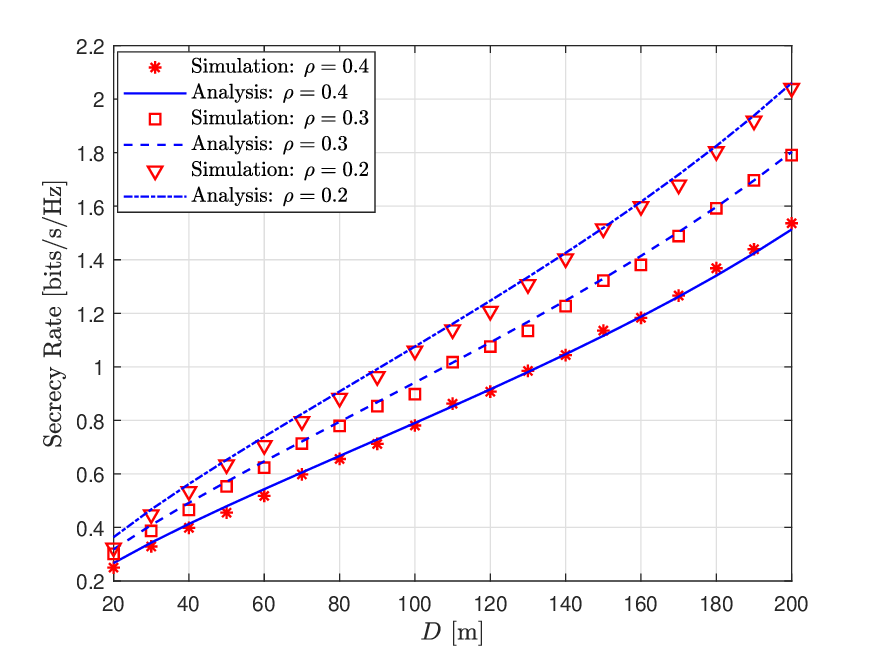}
    \caption{{Secrecy rate versus the  NSZ radius, $D$, for different values of the time allocation factor, $\rho$.  %In this figure, we consider the most detrimental eavesdropper, i.e., the one with the maximum SINR.
    } }
    \label{SRvsD}
    \end{figure}

{Figs.~\ref{SRvsSp} and \ref{SRvsD} show the secrecy rate as a function of the time allocation factor $\rho$ and the NSZ radius $D$, respectively. From these two figures we can observe  that the secrecy rate decreases with increasing $\rho$. This is because a larger portion of time is allocated to energy harvesting, leaving less time for information transmission, which directly reduces the achievable secrecy rate. Moreover, larger values of $D$ improve the secrecy rate due to more effective suppression of nearby interference toward the legitimate UE, which results in higher SINR and hence higher secure rate. This behavior stems from the null-steering mechanism, whereby suppressing radiation toward the reference UE is accompanied by maintaining maximum antenna gain in other directions, which consequently elevates the interference at the eavesdroppers. 
Moreover, the figure shows that the analytical results closely match the simulations in both figures, validating the accuracy of the provided mathematical expression of the secrecy rate in (\ref{secCap2}).}

    \section{Conclusions}
    This paper studied a secure  HAPS-based system with wireless information and power transfer  where FJPT nodes with null-steering capability are deployed to simultaneously enhance secrecy and support energy harvesting for low-power devices. Using stochastic geometry, we derived tractable expressions for the joint rate-energy coverage and the average secrecy rate, which were validated through simulations. A time-switching protocol is used to control the information reception and energy harvesting process at the UE side.  
The results illustrate  important key trade-offs in system design parameters. For instance,  increasing the time-allocation factor improves harvested energy at the expense of communication performance, while enlarging the null-steering zone enhances rate and secrecy but reduces the available harvested energy. Consequently, the joint rate-energy performance exhibits a non-monotonic behavior, indicating the existence of optimal operating points. Overall, the proposed framework demonstrates that properly coordinated null-steering jamming can effectively balance energy harvesting, reliability, and secrecy in HAPS-based networks.
\vspace{-0.3cm}
    \section{Appendix}
    \subsection{Proof of Lemma 1}
Recalling that the interference at the reference UE from the FJPT nodes inside its NSZ and from those outside the NSZ are 
   $I_u^{\mathrm{in}}=\sum_{y_m\in\mathcal{B}(x_o,D)}P_j A_{j}^{\text{null}}g_{ju,m}\ell(d_{ju,m})$ 
and
$I_u^{\mathrm{out}}=\sum_{y_m\in\Phi_j\setminus\mathcal{B}(x_o,D)}P_j A_j g_{ju,m}\ell(d_{ju,m})$. For a given NSZ radius $D$, 
$I_u^{\mathrm{in}}$ and $I_u^{\mathrm{out}}$ are independent. Accordingly, the characteristic functions of $I_u^{in}$, $\Theta_{I_{u}^{\mathrm{in}}}(j\omega)$ is obtained as follows
\begin{eqnarray}
\Theta_{I_{u}^{\mathrm{in}}}(j\omega) &=&\nonumber\\
&&\!\!\!\!\!\!\!\!\!\!\!\!\!\!\!\!\!\!\!\!\!\!\!\!\!\!\!\!\!\!\!\!\!\!\!\!\mathbb{E}_{ \mathcal{B}(0,D),g_{ju}}\Big[\exp\Big(-j\omega \sum_{z_m \in \mathcal{B}(0,D)} P_j A_{j}^{\text{null}}g_{ju,m}\ell(d_{ju,m}\Big)\Big]\nonumber \\
    &&\!\!\!\!\!\!\!\!\!\!\!\!\!\!\!\!\!\!\!\!\!\!\!\!\!\!\!\!\!\!\!\!\!\!\!\!\stackrel{(a)}{=} \mathbb{E}_{ \mathcal{B}(0,D)}\Big[\prod_{z_m \in \mathcal{B}(0,D)} \Big(1 + \frac{j\omega \Omega_j  P_j A_{j}^{\text{null}} C |z_m|^{-{\alpha_j}}}{m_j} \Big)^{-m_j} \Big]\nonumber\\
    &&\!\!\!\!\!\!\!\!\!\!\!\!\!\!\!\!\!\!\!\!\!\!\!\!\!\!\!\!\!\!\!\!\!\!\!\!\stackrel{(b)}{=}\exp\Bigg(-2\pi\lambda_j\int_{0}^{D} \bigg(1 -\nonumber\\
    && \Big(1 + \frac{j\omega \Omega_j  P_j A_{j}^{\text{null}} C z^{-{\alpha_j}}}{m_j} \Big)^{-m_j} \bigg) z\, dz\Bigg).
    \label{a88}
\end{eqnarray}
where $(a)$ follows from the fact that  the small-scale fading, $g_{ju,m}$,  independent gamma random variables, and step (b) is computed by applying the probability generating functional  of the PPP\cite{Integral_hyperfunc.}.  
Let
$\eta \triangleq \frac{j\omega \Omega_j P_j A_{j}^{\text{null}} C}{m_j}$ 
Then, the term in the exponent in (\ref{a88}) can be written as
\begin{equation}
\mathcal{I}
=
\int_{0}^{D}
\left(1-\left(1+\eta z^{-\alpha_j}\right)^{-m_j}\right) z\, dz.
\label{proof_I_def}
\end{equation}
Separating the two terms gives
\begin{equation}
\mathcal{I}
=
\int_{0}^{D} z\, dz
-
\int_{0}^{D}
z\left(1+\eta z^{-\alpha_j}\right)^{-m_j} dz
=
\frac{D^2}{2}
-
\mathcal{J},
\label{proof_I_split}
\end{equation}
where
\begin{equation}
\mathcal{J}
\triangleq
\int_{0}^{D}
z\left(1+\eta z^{-\alpha_j}\right)^{-m_j} dz.
\label{proof_J_def}
\end{equation}
To evaluate \(\mathcal{J}\), we use the change of variable $t=z^{-\alpha_j}$, which implies $z=t^{-1/\alpha_j}, \qquad
dz=-\frac{1}{\alpha_j} t^{-1/\alpha_j-1}dt$. 
Substituting into \eqref{proof_J_def}, we obtain
\begin{equation}
\mathcal{J}
=
\frac{1}{\alpha_j}
\int_{D^{-\alpha_j}}^{\infty}
t^{-\frac{2}{\alpha_j}-1}
(1+\eta t)^{-m_j}
dt.
\label{proof_J_after_t}
\end{equation}
Next, letting \(u=\eta t\), we have \(dt=du/\eta\), and hence
\begin{equation}
\mathcal{J}
=
\frac{\eta^{\frac{2}{\alpha_j}}}{\alpha_j}
\int_{\eta D^{-\alpha_j}}^{\infty}
u^{-\frac{2}{\alpha_j}-1}(1+u)^{-m_j}du.
\label{proof_J_after_u}
\end{equation} 
Now, using the standard Meijer-\(G\) representation
\begin{equation}
(1+u)^{-m_j}
=
\frac{1}{\Gamma(m_j)}
G^{\,1,1}_{1,1}\!\left(
\begin{matrix}
1-m_j\\
0
\end{matrix}
\Big|\,u
\right),
\label{proof_meijerg_basic}
\end{equation}
and applying the integral identity for the product of a power term and a Meijer-\(G\) function, \(\mathcal{J}\) can be expressed in closed form as
\begin{equation}
\mathcal{J}
=
\frac{D^2}{2}\,
\frac{\Gamma\!\left(1-\frac{2}{\alpha_j}\right)}
{\Gamma(m_j)\Gamma\!\left(-\frac{2}{\alpha_j}\right)}
\,G^{\,1,2}_{2,2}\!\left(
\begin{matrix}
1-m_j,\;1+\frac{2}{\alpha_j}\\
0,\;\frac{2}{\alpha_j}
\end{matrix}
\Bigg|\,
-\eta D^{-\alpha_j}
\right).
\label{proof_J_final}
\end{equation}

Substituting \eqref{proof_J_final} into \eqref{proof_I_split} and then using the results into
$\exp\!\left(-2\pi\lambda_j \mathcal{I}\right)$, 
we obtain
\begin{eqnarray}
\Theta_{I_{u}^{\mathrm{in}}}(j\omega)&=&\exp\Bigg( -
     \pi \lambda_j  
     D^2 \Bigg(1-\frac{\Gamma\!\left(1-\frac{2}{\alpha_j}\right)}
{\Gamma(m_j)\Gamma\!\left(-\frac{2}{\alpha_j}\right)}\nonumber\\
&&\!\!\!\!\!\!\!\!\!\!\!\!\!\!\!\!\!\!\!\!\!\!\!\!\!\!\!\!\!\!\!\!\!\!\!\!\!\!\!\!\!\!\!\!
\times\,G^{\,1,2}_{2,2}\!\left(
\begin{matrix}
1-m_j,\;1+\frac{2}{\alpha_j}\\
0,\;\frac{2}{\alpha_j}
\end{matrix}
\bigg|\, 
     \frac{-j\omega C P_j A_j^{\text{null}}D^{-\alpha_j}\Omega_j}{m_j }
\right)\Bigg)\Bigg),
\label{proof17}
\end{eqnarray}
which is the result in (\ref{CF1}). To derive  $\Theta_{I_{u}^{\mathrm{out}}}(j\omega)$, it is first expressed as
\begin{eqnarray}
\Theta_{I_{u}^{\mathrm{out}}}(j\omega) &=&\nonumber\\
&&\!\!\!\!\!\!\!\!\!\!\!\!\!\!\!\!\!\!\!\!\!\!\!\!\!\!\!\!\!\!\!\!\!\!\!\!\mathbb{E}_{\Phi_j,g_{ju},A_j}\Big[\exp\Big(-j\omega \sum_{z_m \in \Phi_j\setminus \mathcal{B}(0,D)} P_j A_{j}g_{ju,m}\ell(d_{ju,m}\Big)\Big]\nonumber \\
    &&\!\!\!\!\!\!\!\!\!\!\!\!\!\!\!\!\!\!\!\!\!\!\!\!\!\!\!\!\!\!\!\!\!\!\!\!= \mathbb{E}_{\Phi_j,A_j}\Big[\prod_{z_m \in \Phi_j\setminus\mathcal{B}(0,D)} \!\!\!\!\!\!\!\Big(1 + \frac{j\omega \Omega_j  P_j A_{j} C |z_m|^{-{\alpha_j}}}{m_j} \Big)^{-m_j} \Big]\nonumber\\
    &&\!\!\!\!\!\!\!\!\!\!\!\!\!\!\!\!\!\!\!\!\!\!\!\!\!\!\!\!\!\!\!\!\!\!\!\!=\mathbb{E}_{A_j}\Bigg[\exp\Bigg(-2\pi\lambda_j\int_{D}^{R_s} \bigg(1 -\nonumber\\
    && \!\!\!\!\!\!\!\!\!\!\Big(1 + \frac{j\omega \Omega_j  P_j A_{j} C z^{-{\alpha_j}}}{m_j} \Big)^{-m_j} \bigg) z\, dz\Bigg)\Bigg]\nonumber\\
    &&\!\!\!\!\!\!\!\!\!\!\!\!\!\!\!\!\!\!\!\!\!\!\!\!\!\!\!\!\!\!\!\!\!\!\!\!\stackrel{(a)}{=} \exp\Bigg(-2\pi\lambda_j\sum_{q=1}^2p_q\int_{D}^{R_s} \bigg(1 -\nonumber\\
    && \!\!\!\!\!\!\!\!\!\!\Big(1 + \frac{j\omega \Omega_j  P_j A_{q} C z^{-{\alpha_j}}}{m_j} \Big)^{-m_j} \bigg) z\, dz\Bigg)
    \label{a89}
\end{eqnarray}
where step (a) follows from averaging over the random antenna gain $A_j$ using its model in (\ref{eq:null_model}). Then, applying the steps (\ref{proof_I_def})-(\ref{proof17}) for (\ref{a89}), we get the results in (\ref{CF2}) which complete the proof. \hfill $\blacksquare$
\bibliographystyle{IEEEtran}
\bibliography{Ref}
\end{document}